\documentclass[12pt]{iopart}

\usepackage[dvips]{graphicx}
\usepackage{epsfig}
\usepackage{psfig,epsfig,pstricks}
\usepackage{placeins}
\usepackage{xspace}

\newcommand{\ie}{{\it i.e.}\xspace}
\newcommand{\eg}{{\it e.g.}\xspace}
\newcommand{\G}{\mathcal{G}}

\newcommand{\smax}{s_{\mathrm{max}}}
\newcommand{\smin}{s_{\mathrm{min}}}

\newcommand{\ave}[1]{\left\langle#1 \right\rangle}

\newcommand{\elabel}[1]{\label{#1}}

\newcommand{\flabel}[1]{\label{#1}}

\newcommand{\tlabel}[1]{\label{#1}}

\usepackage{subfigure}

\begin{document}

\title{Universality of rain event size distributions}
 
\author{O. Peters$^{1,2}$, A. Deluca$^3$, A. Corral$^3$, J. D. Neelin$^2$, C. E. Holloway$^4$}
\ead{ole@santafe.edu}
\address{
$^1$Dept. of Mathematics and Grantham Institute for Climate Change, Imperial College London, 180 Queen's Gate, London SW7 2AZ, UK.\\
$^2$Dept. of Atmospheric and Oceanic Sciences and Institute for Geophysics and Planetary Physics,
University of California, Los Angeles, 405 Hilgard Ave., Los Angeles,
California 90095-1565, USA\\
$^3$Centre de Recerca Matem\`atica, Campus de Bellaterra, Edifici C, 08193 Bellaterra (Barcelona), Spain\\
$^4$Dept. of Meteorology, University of Reading, Earley Gate, PO Box 243, Reading RG6 6BB, UK
}
\date{\today}

\begin{abstract}

We compare rain event size distributions derived from measurements in
climatically different regions, which we find to be well approximated
by power laws of similar exponents over broad ranges. Differences can
be seen in the large-scale cutoffs of the distributions. Event
duration distributions suggest that the scale-free aspects are related
to the absence of characteristic scales in the meteorological
mesoscale.
\end{abstract}




\vspace{2pc}
\submitto{J. Stat. Mech.}
\maketitle

\section{Introduction}
Atmospheric convection and precipitation have been hypothesised to be
a real-world realization of self-organized criticality (SOC). This idea
is supported by observations of avalanche-like rainfall events
\cite{AndradeSchellnhuberClaussen1998,PetersHertleinChristensen2002}
and by the nature of the transition to convection in the atmosphere
\cite{PetersNeelin2006,NeelinPetersHales2009}. Many questions remain
open, however, as summarized below. Here we ask whether the
observation of scale-free avalanche size distributions is reproducible
using data from different locations and whether the associated fitted
exponents show any sign of universality.

Many atmospheric processes are characterized by long-range spatial and
temporal correlation, and by corresponding structure on a wide range
of scales. There are two complementary explanations why this is so,
and both are valid in their respective regimes: structure on many
scales can be the result of different processes producing {\it many}
characteristic scales \cite{Klein2010,BodenschatzETAL2010}; it can
also be the result of an {\it absence} of characteristic scales over
some range, such that all intermediate scales are equally significant
\cite{Barenblatt1996}. The latter perspective is relevant, for
instance, in critical phenomena and in the inertial subrange of fully
developed turbulence.

Processes relevant for precipitation are associated with many different
characteristic time and spatial scales, see \eg
Ref.~\cite{BodenschatzETAL2010}.
The list of these scales has a gap, however, from a few km (a few
minutes) to 1,000~km (a few days), spanning the so-called mesoscale,
and it is in this gap that the following arguments are most likely to
be relevant.

The atmosphere is slowly driven by incident solar radiation, about
half of which is absorbed by the planet's surface, heating and
moistening the atmospheric boundary layer; combined with radiative
cooling at the top of the troposphere this creates an instability. This
instability drives convection, which in the simplest case is dry. More
frequently, however, moisture and precipitation play a key role. Water
condenses in moist rising air, heating the environment and reinforcing
the rising motion, and often, the result of this process is
rainfall. The statistics of rainfall thus contain information about
the process of convection and the decay towards stability in the
troposphere. A common situation is conditional instability, where
saturated air is convectively unstable, whereas dry air is
stable. Under-saturated air masses then become unstable to convection
if lifted by a certain amount, meaning that relatively small
perturbations can trigger large responses.

Since driving processes are generally slow compared to convection, it
has been argued that the system as a whole should typically be in a
far-from equilibrium statistically stationary state close to the onset
of instability. In the parlance of the field this idealized state,
where drive and dissipation are in balance, is referred to as
``Quasi-Equilibrium'' (QE) \cite{ArakawaSchubert1974}. In
Ref.~\cite{PetersNeelin2006}, using satellite data over tropical
oceans, it was found that departures from the point of QE into the
unstable regime can be described as triggering a phase transition
whereby large parts of the troposphere enter into a convectively
active phase. Assuming that the phase transition is continuous, the
attractive QE state would be a case of SOC -- a critical point of a
continuous phase transition acting as an attractor in the phase space
of a system \cite{TangBak1988,DickmanVespignaniZapperi1998}.

The link between SOC and precipitation processes has also been made by
investigating event-size distributions in a study using data from a
mid-latitude location \cite{PetersHertleinChristensen2002}. Both the
tropical data in Ref.~\cite{PetersNeelin2006} and the mid-latitude
data in Ref.~\cite{PetersHertleinChristensen2002} support some notion
of SOC in precipitation processes, but the climatologies in these
regions are very different. Rainfall in the mid-latitudes is often
generated in frontal systems, whereas in the tropics, much of the
precipitation is convective, supporting high rain rates. It is not
{\it a priori} clear whether these differences are relevant to the SOC
analogy, or whether they are outweighed by the robust similarities
between the systems. For instance, drive and dissipation time scales
are well separated also in the mid-latitudes. In time series from
Sweden the average duration of precipitation events was found to be
three orders of magnitude smaller than the average duration of dry
spells \cite{OlssonNiemczynowiczBerndtsson1993}. It is therefore
desirable to compare identical observables from different locations.

Scale-free event size distributions suggest long-range correlation in
the system, which in turn hints at a continuous transition to
precipitation. Similar effects, however, can also result directly from
a complex flow field, as was shown in simulations using randomized
vortices and passive tracers \cite{Dickman2003}. Since the fluid
dynamics is complex enough to generate apparent long-range
correlation, and it is difficult from direct observation to judge
whether the transition is continuous, we cannot rule out a
discontinuous jump.

This uncertainty is mirrored in parameterizations of convection. The
spatial resolution of general circulation models is limited by
constraints in computing power to about 100~km in the
horizontal. Dynamically there is nothing special about this scale, and
the approach in climate modeling for representing physical processes
whose relevant spatial scales are smaller is to describe their
phenomenology in parameterizations.  Parameterizations of convection
and precipitation processes often contain both continuous and
discontinuous elements. For instance, the intensity of convection and
precipitation typically depends continuously on a measure of
convective plume buoyancy (such as convective available potential
energy) and water vapor content
\cite{ArakawaSchubert1974,BettsMiller1986}, but sometimes a
discontinuous threshold condition is introduced to decide whether
convection occurs at all \cite{NeelinETAL2008}.

\section{Data sets}
We study rain data from all 10 available sites of the Atmospheric
Radiation Measurement (ARM) Program, see www.arm.gov, over periods
from about 8 months to 4 years, see Table~1. Precipitation rates were
recorded at one-minute resolution, with an optical rain gauge, Model
ORG-815-DA MiniOrg (Optical Scientific, Inc.)
\cite{ARMhandbooks}. Data were corrected using the ARM Data Quality
Reports \cite{ARMquality}, and rates below 0.2~mm/h were treated as
zero measurements, as recommended by the ARM Handbook
\cite{ARMhandbooks}, see \fref{intensities}.

The measurements are from climatically different regions using a
standardized technique, making them ideal for our purpose. Three sites
are located in the Tropical Western Pacific (Manus, Nauru and Darwin),
known for strong convective activity. Niamey is subject to strong
monsoons, with a pronounced dry season. Heselbach is a mid-latitude
site with an anomalously large amount of rainfall due to orographic
effects. Rainfall in Shouxian is mostly convective in the summer
months, which constitute most of the data set. Graciosa Island in the
Azores archipelago is a sub-tropical site, chosen for the ARM program
to study precipitation in low clouds of the marine boundary layer. 

Three data are less straight-forward: The Point Reyes measurements
specifically target Marine Stratus clouds, which dominate the
measurement period and are known to produce drizzle in warm-cloud
conditions (without ice phase). Unfortunately the measurements only
cover six months, and it is unclear whether observed differences are
due to the different physics or to the small sample size.
The Southern Great Plains (SGP) measurements suffer from a malfunction
that led to apparent rain rates of about 0.1~mm/h over much of the
observation period. The problem seems to be present in most other data
sets but is far less pronounced there, see \fref{intensities}.
Measurements at
temperatures below $3^\circ$C were discarded as these can contain snow
from which it is difficult to infer equivalent rates of liquid water
precipitation. The North Slope of Alaska (NSA) data set contains
mostly snow; it is included only for completeness.

None of the data sets showed significant seasonal variations in the
scaling exponents. In the Point Reyes, SGP and NSA data we found
slight variations but could not convince ourselves that these were
significant. Data from all seasons are used.

\begin{figure}
\centering
\includegraphics*[height=8.5cm]{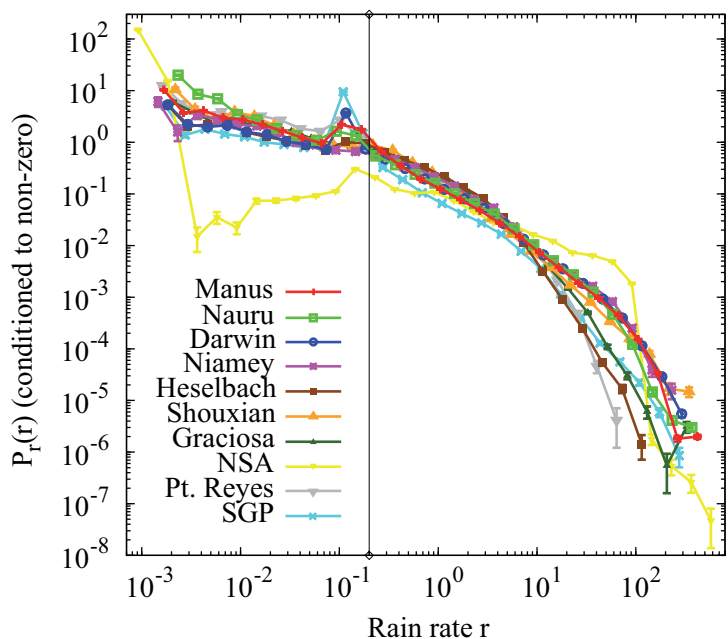}

\caption{Probability (relative frequency) density of precipitation
  rate, $r$ in mm/h. The vertical line indicates the lower intensity cutoff at
  0.2~mm/h. Smaller rain rates are treated as zero. The peak around
  0.1~mm/h, most ponounced in the Southern Great Plains data, is due
  to a malfunction of the instrument. The Alaska data set contains
  mostly snow and is included only for
  completeness. \flabel{intensities}}
\end{figure}

\begin{table}[ht]
\caption{Observation sites with corresponding time periods, number of observed 
   precipitation events $N$, estimated annual precipitation in mm, and location.}
\begin{tabular}{ c c c c c c}
\br
\footnotesize{Site} & \footnotesize{From} & \footnotesize{Until} &  \footnotesize{$N$} & \footnotesize{Precip./yr} &  \footnotesize{Location} \\
\mr
 \footnotesize{Manus Island,}& \footnotesize{02/15/2005} & \footnotesize{08/27/2009} & \footnotesize{11981} &  \footnotesize{5883.29 } & \footnotesize{2.116$^\circ$ S, 147.425$^\circ$ E}\\ 
\footnotesize{Papua New Guinea}&&&& &\\ 
\footnotesize{Nauru Island,} & \footnotesize{02/15/2005} & \footnotesize{08/27/2009 } & \footnotesize{5134} & \footnotesize{1860.87  } 
& \footnotesize{0.521$^\circ$ S, 166.916$^\circ$ E}\\
  \footnotesize{Republic of Nauru}&&&&&\\ 
\footnotesize{Darwin, Australia}		 & \footnotesize{02/15/2005} & \footnotesize{08/27/2009}  & \footnotesize{2883} & \footnotesize{1517.09  } 
& \footnotesize{12.425$^\circ$ S, 130.892$^\circ$ E} \\ 
 \footnotesize{Niamey, Niger}									 & \footnotesize{12/26/2005} & \footnotesize{12/08/2006}  & \footnotesize{262} & \footnotesize{608.37  }
& \footnotesize{13.522$^\circ$ N, 2.632$^\circ$ E}\\ 
 \footnotesize{Heselbach, Germany}							 & \footnotesize{04/01/2007} & \footnotesize{01/01/2008}  & \footnotesize{2439} &
 \footnotesize{2187.85  } 
& \footnotesize{48.450$^\circ$ N, 8.397$^\circ$ E}\\ 
\footnotesize{Shouxian, China}								 & \footnotesize{05/09/2008} & \footnotesize{12/28/2008}  & \footnotesize{480} & \footnotesize{1221.20  } 
& \footnotesize{32.558$^\circ$ N, 116.482$^\circ$ E}\\
\footnotesize{Graciosa Island, Azores}								 & \footnotesize{04/14/2009} & \footnotesize{07/10/2010}  & \footnotesize{3066} & \footnotesize{702.35  } 
& \footnotesize{39.091$^\circ$ N, 28.029$^\circ$ E}\\
\footnotesize{NSA, USA}								 & \footnotesize{04/01/2001} & \footnotesize{10/13/2003}  & \footnotesize{9097} & \footnotesize{23516.16  } 
& \footnotesize{71.323$^\circ$ N, 156.616$^\circ$ E}\\
\footnotesize{Point Reyes, USA}								 & \footnotesize{02/01/2005} & \footnotesize{09/15/2005}  & \footnotesize{579} & \footnotesize{ 797.85  } 
& \footnotesize{38.091$^\circ$ N, 122.957$^\circ$ E}\\
\footnotesize{SGP, USA}		 & \footnotesize{11/06/2007} & \footnotesize{08/24/2009}  & \footnotesize{1624} & \footnotesize{968.95 } 
& \footnotesize{36.605$^\circ$ S, 97.485$^\circ$ E} \\ 
\br
 \end{tabular}
 \label{table: sites} 
\end{table} 

\section{Event sizes}
The data used here are (0+1)-dimensional time series, whereas the
atmosphere is a (3+1)-dimensional system. We leave the question
unanswered which spatial dimensions are most relevant -- the system
becomes vertically unstable, but it also communicates in the two
horizontal dimensions through various processes
\cite{NeelinPetersHales2009}.

Following Ref.~\cite{PetersHertleinChristensen2002}, we define an
event as a sequence of non-zero measurements of the rain rate, see
inset in \fref{pdfevents}. The event size $s$ is the rain rate,
$r(t)$, integrated over the event, $s=\int_{\mathrm{event}} dt~
r(t)$. The dimension of this object is $\left[s\right]=$mm, specifying
the depth of the layer of water left on the ground during the
event. One mm corresponds to an energy density of some 2500 kJ/m$^2$
released latent heat of condensation. If the rain rate were known over
the area covered by the event, then the event size could be defined
precisely as the energy released during one event. Since spatial
information is not available, it is ignored in our study. 

\begin{figure}
\centering
\includegraphics*[height=8.5cm]{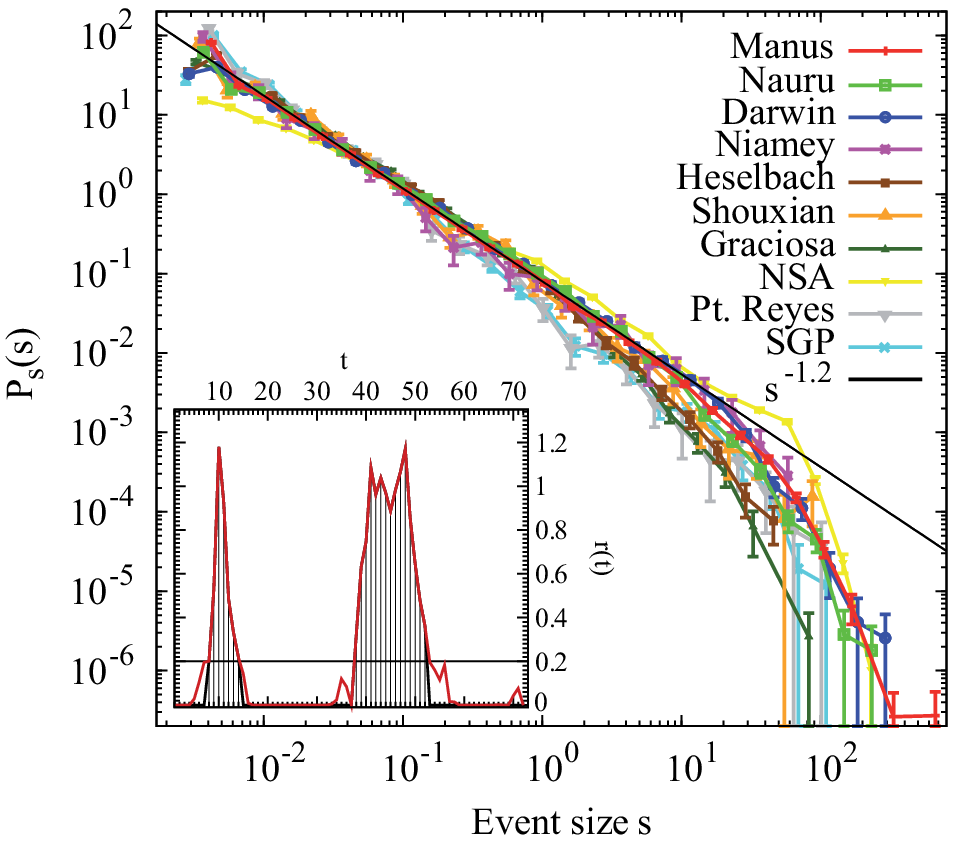}
\caption{Probability densities of event sizes, $s$ in mm, and a
  power-law fit (black straight line).  \\ \underline{ Inset}:
  Precipitation rates from Niamey, including two rain events lasting 7
  and 15 minutes respectively. Interpreting reported rain rates of
  less than 0.2 mm/h as zero, the shaded areas are the corresponding
  event sizes.  \flabel{pdfevents}  }  
\end{figure}

For each data set, the probability density function $P_s(s)$ in a
particular size interval $[s, s + \Delta s)$ is estimated as
  $P_s(s)\approx {n(s)}/({N\Delta s})$, where $n(s)$ is the number of
  events in the interval and $N$ the total number of events.  We use
  $(s+\Delta s)/s = 10^{1/5} \approx 1.58$, \ie 5 bins per order of
  magnitude in $s$.  Standard errors are shown, for $P_s(s)$: assuming
  Poissonian statistics, the error in $n(s)$ is approximated by
  $\sqrt{n(s)}$.

\section{SOC scaling}
Studies of simple SOC models that approach the critical point of a
continuous phase transition focus on avalanche size distributions,
which we liken to rain event sizes. Critical exponents are derived
from finite-size scaling, that is, the scaling of observables
with system size (as opposed to critical scaling, the scaling of
observables with the distance from criticality).
In SOC models, moments of the
avalanche size distribution scale with system size $L$ like
\begin{equation}
\ave{s^k}\propto L^{D(1+k-\tau_s)}  \mathrm{\hspace{1mm} for \hspace{1mm}} k > \tau_s-1,
\elabel{fss}
\end{equation}
defining the exponent $D$, sometimes called the avalanche dimension,
and the exponent $\tau_s$, which we call the avalanche size exponent.
\Eref{fss} is consistent with probability density functions $P_s(s)$
of the form
\begin{equation}
P_s(s)=s^{-\tau_s} \G_s(s/s_\xi) \mathrm{ \hspace{1mm} for  \hspace{1mm}} s>s_l
\elabel{density}
\end{equation}
where $s_\xi=L^D$, and the scaling function $\G_s(s/s_\xi)$ falls off
very fast for large arguments, $s/s_\xi>1$, and is constant for small
arguments, $s/s_\xi\ll 1$, down to a lower cutoff, $s=s_l$, where
non-universal microscopic effects ({\it e.g.} discreteness of the
system) become important.

Assuming that we have observations from an SOC system, and that a
significant part of the observed avalanche sizes are in the region
$s_l <s\ll s_\xi$, we expect to find a range of scales where the power
law
\begin{equation}
P_s(s) = \G_s(0) s^{-\tau_s}
\elabel{power_law}
\end{equation}
holds. Under sufficiently slow drive the exponent $\tau_s$ is believed
to be robust in SOC models
\cite{AlavaETAL2008,PruessnerPeters2008}. We infer event-size
distributions like in Ref.~\cite{PetersHertleinChristensen2002} from
measurements in different locations and compare values for the
apparent avalanche size exponent $\tau_s$. As a first step to assess
the validity of \eref{power_law} we produce log-log plots of $P_s(s)$
vs. $s$ and look for a linear regime, \fref{pdfevents}. Since the
study of critical phenomena is a study of limits that cannot be
reached in physical systems, the field is notorious for debates
regarding the significance of experimental work, which is especially
true for SOC. While an element of interpretation necessarily remains,
we devise methods to maximize the objectivity of our analysis.

In our data sets, time series of rain rates from different locations,
we interpret the upper limit $s_\xi$ of the scale-free range as an
effective system size. We cannot control this size; nonetheless the
scaling hypothesis, \eref{density}, can be tested using appropriate
moment ratios \cite{RossoLeDoussalWiese2009}. For instance,
$s_{\xi}\propto {\ave{s^2}}/{\ave{s}}$, provided $s_l\ll
s_\xi$. Hence, to account for changes in effective system sizes the
$s$-axis in \fref{pdfevents} can be rescaled to $s
{\ave{s}}/{\ave{s^2}}$, see \fref{scalingfunction2_a}. This collapses
the loci of the large-scale cutoffs. Plotting $P_s(s)s^{\tau_s}$
against this rescaled variable produces \fref{scalingfunction2_b} of
the scaling function $\G_s(s/(a s_\xi))$, where $a$ is the
proportionality constant relating $s_{\xi}$ to the moment ratio. This
has the advantage of reducing the logarithmic vertical range, which
makes it possible to see differences in the distributions that would
otherwise be concealed visually. \Fref{scalingfunction2_a} covers 9
orders of magnitude vertically, whereas \fref{scalingfunction2_b}
covers little more than 2.

\begin{figure}
\centering
\subfigure[]{
\flabel{scalingfunction2_a}
\includegraphics*[height=0.40\textwidth]{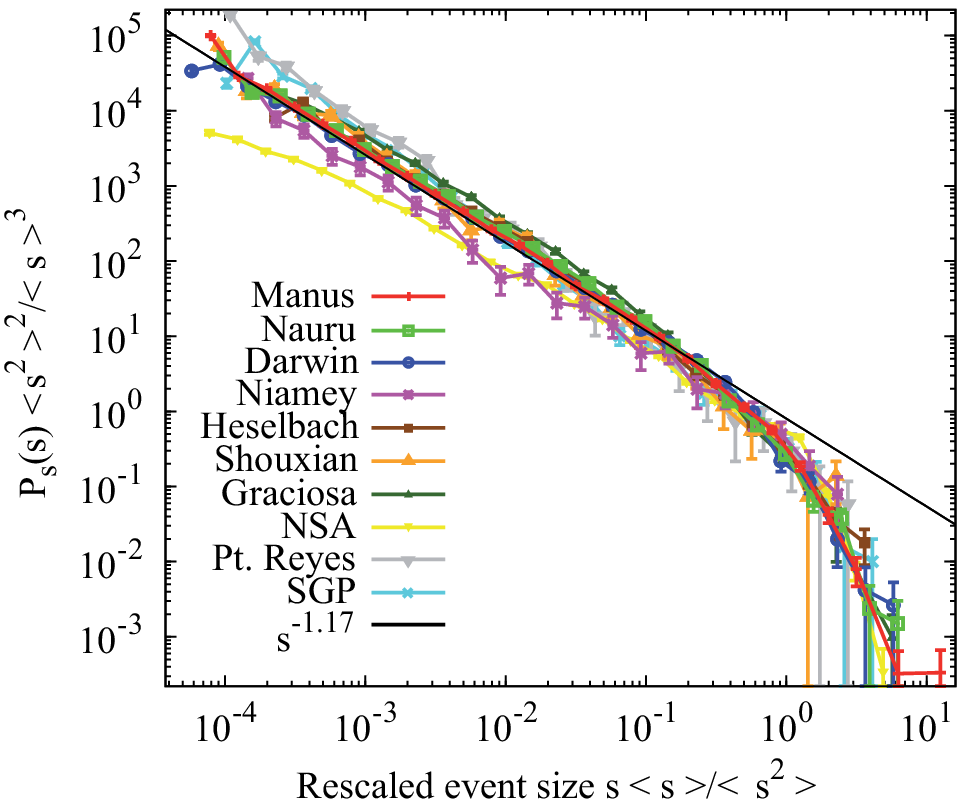}
}
  		 \hspace{0.01cm}
\subfigure[]{
\flabel{scalingfunction2_b}
\includegraphics*[height=0.40\textwidth]{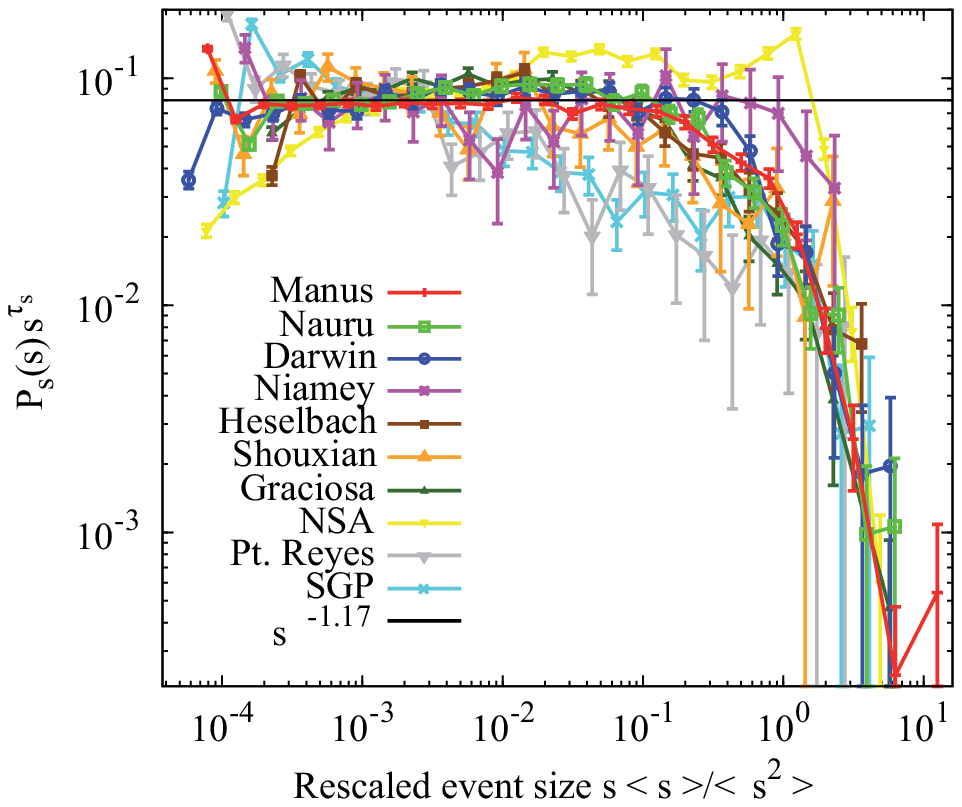}}
 \caption{ (a) Event size distributions rescaled with the moment ratio
   $\frac{\ave{s^2}}{\ave{s}}$. 
   (b) Inferred scaling function $\G_s$,
   using $\tau_s=1.17$ for all data sets. By far the largest
   deviations from a common scaling function are observed for the
   unreliable data sets, Alaska (NSA) and Southern Great Plains (SGP).
 } 
\flabel{scalingfunction2}
 \end{figure}

\section{Exponent estimation and goodness of fit}

For a detailed discussion, see \ref{Fitting}.  We apply a form of
Kolmogorov-Smirnov (KS) test \cite{PressETAL2002} similar to that in
Ref. \cite{ClausetShaliziNewman2009}.  First, a fitting range $[\smin,
  \smax]$ is selected. In this range the maximum-likelihood value for
$\tau_s$ in \eref{power_law} is found. Next, the maximum difference
between the empirical cumulative distribution in this range and the
cumulative distribution corresponding to the best-fit power law is
found. The same measure is applied to synthetic samples of data (each
with the same number of instances), generated from the best-fit
power-law distribution. This yields the ``p''-value, \ie the fraction
of samples generated from the tested model (the best-fit power law)
where at least such a difference is observed. We stress that each
synthetic data set is compared to its own maximum-likelihood power-law
distribution, \ie an exponent has to be fitted for each sample, so
that no bias be introduced.

We keep a record of the triplet $(\smin, \smax, \tau_s)$ if the
$p-$value is greater than 10\% (our arbitrarily chosen threshold).
After trying all possible fitting ranges with $\smin$ and $\smax$
increasing by factors of $10^{0.01}$, we select the
triplet which maximizes the number $\bar N$ of data between $\smin$
and $\smax$.

The distributions in \fref{pdfevents} are visually compatible with a
power law (black straight line) over most of their ranges. The
procedure consisting of maximum-likelihood estimation plus a
goodness-of-fit test confirms this result: over ranges between 2 and 4
orders of magnitude, all data sets are consistent with a power-law
distribution and the estimates of the apparent exponents are in
agreement with the hypothesis of a single exponent $\tau_s=1.17(3)$,
brackets indicating the uncertainty in the last digit, except for the
three problematic data sets from Point Reyes, the Southern Great
Plains and Alaska. The complete results are collected in
Table~\ref{table: maxnumber10}. While the best-fit exponents in this
table are surprisingly similar (given the climatic differences between
the measuring sites), the error estimates are unrealistically
small. Taking the statistical results literally, we would have to
conclude that the exponents are very similar but mutually incompatible
(\eg $\tau_s{^\mathrm{Manus}}=1.18(1)$ and
$\tau_s{^\mathrm{Nauru}}=1.14(1)$) suggesting that $\tau_s$ is not
universal. On physical grounds we do not believe this conclusion
because systematic errors arising from the measurement process, the
introduction of the sensitivity threshold, binning during data
recording {\it etc}., are likely to be much larger than the purely
statistical errors quoted here. For example,
Ref.~\cite{PetersHertleinChristensen2002} used a different type of
measurement with a smaller sensitivity threshold and led to a best
estimate for the exponent of 1.36. Furthermore, the apparent exponent
can only be seen as a rough estimate of any true underlying
exponent. We tested that, fixing $\tau_s=1.17$, all data sets yield $p
> 10 \%$ over a range larger than two and a half orders of magnitude,
except for the three problematic data sets. A two-sample
Kolmogorov-Smirnov test for all pairs of datasets further confirms the
similarity of the distributions for the different sites,
\ref{Two-sample}.

In ~\fref{surface_a} we show a color plot of all triplets $(\smin,
\smax, \tau_s)$, corresponding to the Manus dataset. There
is a large plateau where $\tau_s\approx 1.17$, indicating that this
value is the best estimate for many intervals. \Fref{surface_b} is
an analogous plot for the $p-$value, showing that the goodness of the
fit is best in the region of the plateau.

\begin{figure}
\centering 
\subfigure[]{\flabel{surface_a}\includegraphics*[height=0.40\textwidth]{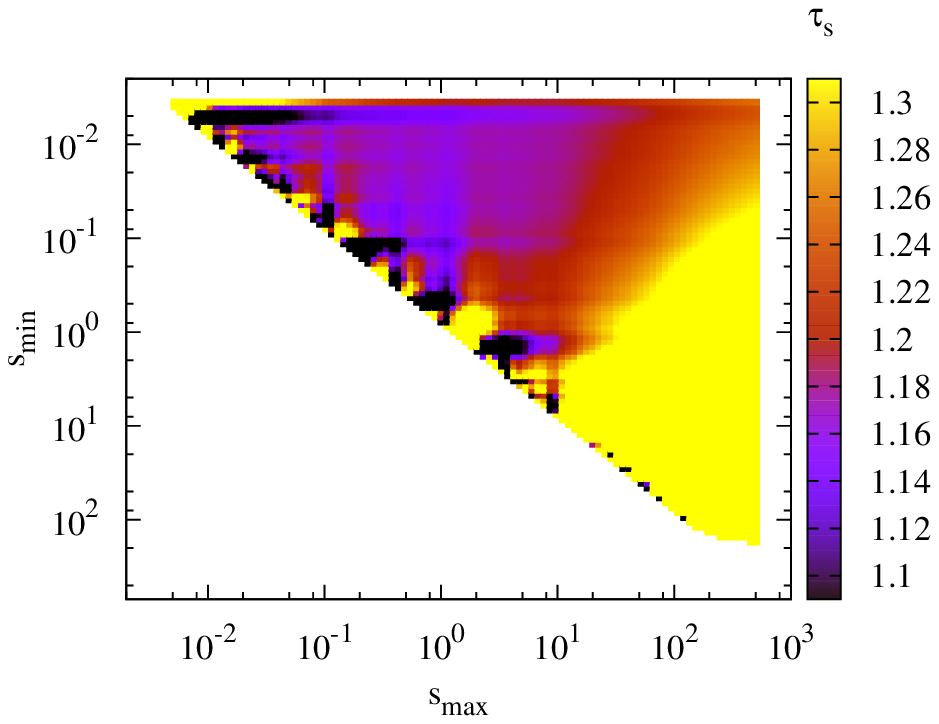}}
\hspace{0.01cm} 
\subfigure[]{\flabel{surface_b}\includegraphics*[height=0.40\textwidth]{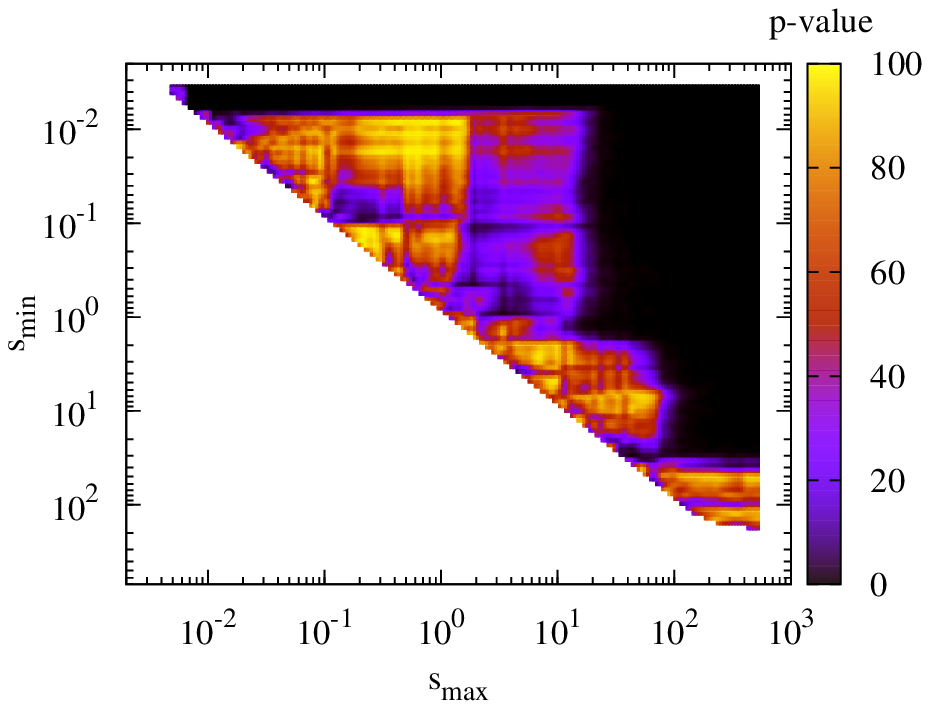}}
\caption{ (a) Color map showing the best-fit value for the exponent
   $\tau_s$ for all pairs of $\smin$ and $\smax$, (lower and upper
   ends of the chosen fitting range in mm) for the Manus dataset.  The
   large plateau corresponds to $\tau_s\approx 1.17$. (b) Analogous
   plot for the $p-$value.  } \flabel{surface}
 \end{figure}  

\begin{table}
\caption{Avalanche size exponent $\tau_s$ for all sites (last column).
  Lower and upper end of fitting range (in mm), logarithmic range
  $\smax/\smin$, number of events $N$, number of events in fitting
  range $\bar N$, and a moment ratio proportional to the cutoff
  $s_\xi$ are shown. Brackets () denote errors in the last digit,
  determined by jackknife \cite{Efron1982}.
\label{table: maxnumber10}}	
 \begin{indented}
\item[] 
\begin{tabular}{l r r r r r c c} 
\br
\textbf{Site}& $s_{\mathrm{min}}$  & $s_{\mathrm{max}}$  & ${s_{\mathrm{max}}}/{s_{\mathrm{min}}}$ & $N$ & $\overline{N}$   & \small{$\ave{s^2}/\ave{s}(er)$} & $\tau_s(er)$ \\ 
\mr
\footnotesize{Manus} &        0.0069 &      18.7 &      2719. &    11981 &     9320 & 53.(1) &    1.19(1)\\
\footnotesize{Nauru}  &       0.0066 &      4.7 &      704. &     5134 &     3996 & 37.(1) &   1.14(1)\\
\footnotesize{Darwin}  &      0.0067 &      21.6 &      3230. &     2883 &     2410 & 50.(1)&   1.16(1)\\
\footnotesize{Niamey}  &      0.0041 &      55.0 &      13500. &      262 &      232 & 25.(2)& 1.19(3)\\ 
\footnotesize{Heselbach}  &   0.0072 &      1.4 &      195. &     2439 &     1764 & 13.(1)&   1.18(2)\\ 
\footnotesize{Shouxian}  &    0.0037 &      2.5 &      677. &      480 &      406  & 39.(2)&    1.19(3)\\ 
\footnotesize{Graciosa}  &    0.0069 &      1.0 &   148. &     3066 &     2260 & 14.4(3) &  1.16(1)\\
\footnotesize{NSA}  &   0.0205      &   5.9   &    288.  &   9097   & 6030 & 47.(1) &   1.01(1)\\
\footnotesize{Pt. Reyes}  &   0.0062 &      66.7 &   10796. &      579  &      427 &  37.(2)&  1.40(2)\\
\footnotesize{SGP}  &         0.0062 &     58.8 &       9463. &  1624 &   1196 &   27.(1) &  1.40(2) \\  
 \br                        
 \end{tabular}
\end{indented}
\end{table}

Climatic differences between regions are scarcely detectable in event
size distributions, which may be surprising on the grounds of
climatological considerations. However, the cutoff $s_\xi$,
representing the capacity of the climatic region around a measuring
site to generate rain events, changes significantly from region to
region, confirming meteorological intuition. This is difficult to see
in the logarithmic scales of \fref{pdfevents} but is easily extracted
from the moments of the distributions, Table~\ref{table: maxnumber10}.
Thus, the smallest cutoff (and likely maximum event size) in the ARM
data is found in Heselbach (mid-latitudes), whereas the largest is in
Manus (Western Pacific warm pool). We note that $\ave{s^2}/\ave{s}$ is
only proportional to the actual cutoff $s_\xi$. Assuming a box
function for the scaling function and using the value $\tau_s=1.17$,
we can estimate the proportionality constant and find $s_\xi \approx
2.2 \ave{s^2}/\ave{s}$. With this estimate, none of the fitting ranges
extends beyond the cutoff.

\section{Dry spells}
The durations of precipitation-free intervals have also been reported
to follow an approximate power law
\cite{PetersHertleinChristensen2002}. We therefore repeat for
dry-spell durations the same analysis as for the event
sizes. \Fref{fdryspells_a} shows the distributions, with a collapse
corresponding to \fref{scalingfunction2_b} in \fref{fdryspells_b}. We
notice the different strengths of the diurnal cycle, here visible as a
relative peak near 1 day dry spell duration. Exponents fitted to the
distributions are similar, see \tref{tdryspells}.

 \begin{figure}
 \centering
       \subfigure[]{\flabel{fdryspells_a}\includegraphics*[height=0.40\textwidth]{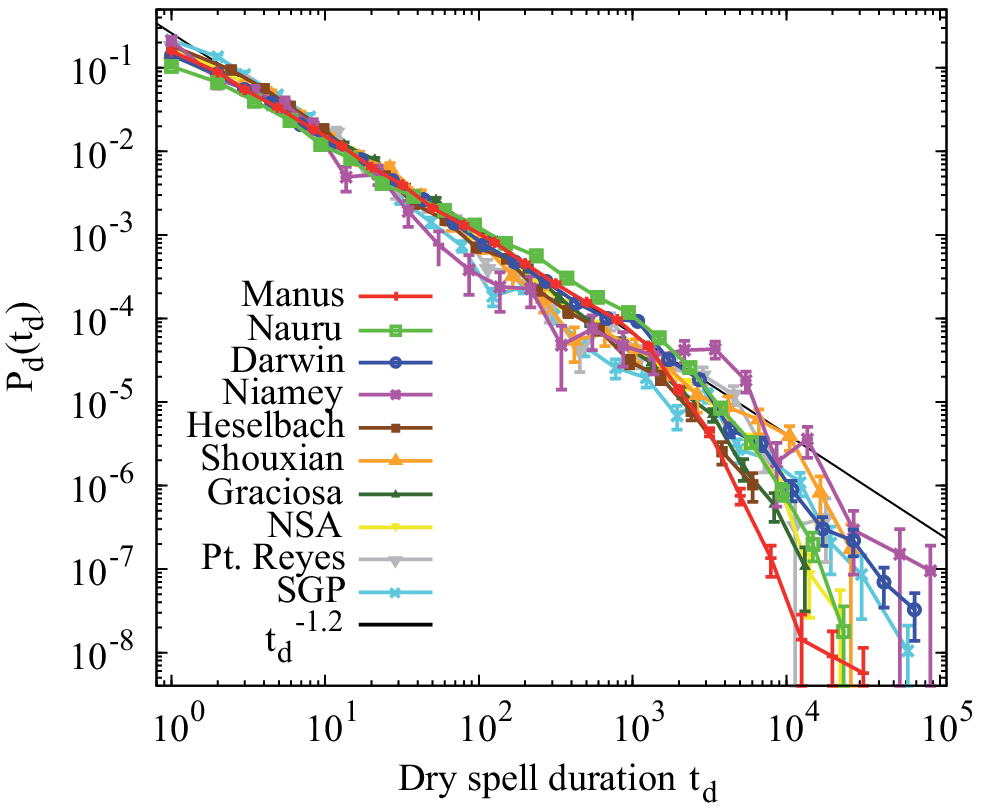}}
\hspace{0.01cm}
\subfigure[]{\flabel{fdryspells_b}\includegraphics*[height=0.412\textwidth]{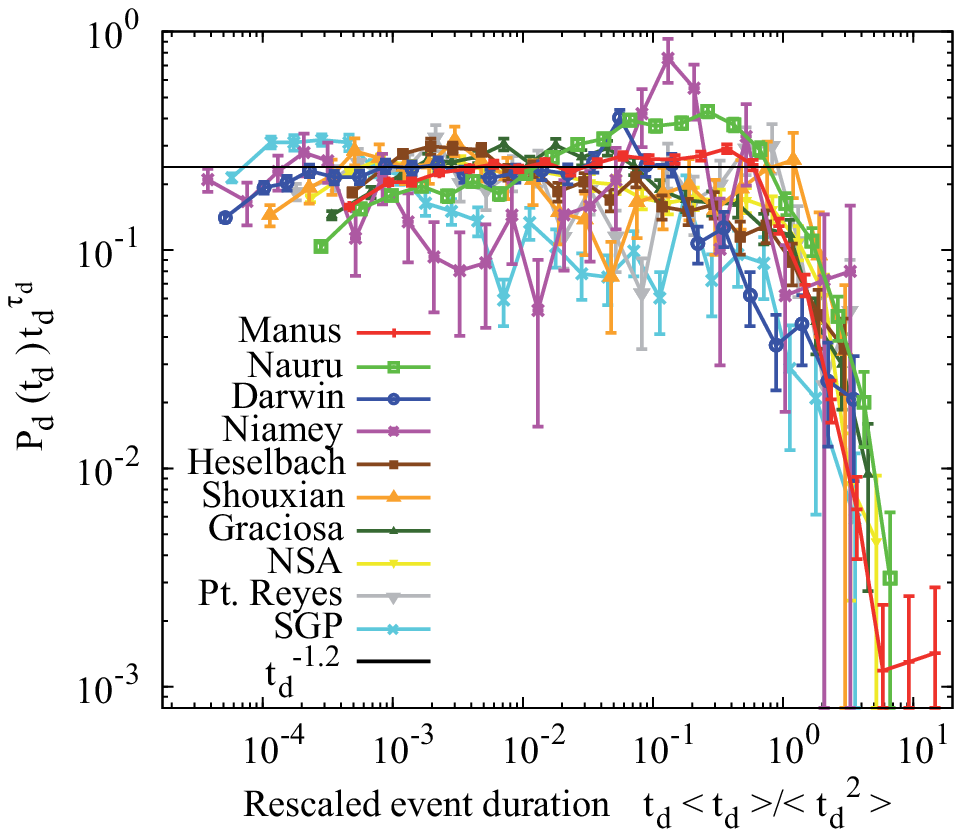}}
 \caption{ (a) Probability densities for dry spell durations (in
   min). The diurnal cycle is most pronounced in Niamey, otherwise the
   distributions are similar. (b) Distributions collapsed onto their
   scaling function, similar to \fref{scalingfunction2_b}.  }
 \flabel{fdryspells}
 \end{figure}  

\begin{table}[!hb]
\caption{ Dry spell exponent (last column). Lower and upper end of
  fitting range (in min), logarithmic range
  ${t_{d}}_{\mathrm{max}}/{t_{d}}_{\mathrm{min}}$, number of dry
  spells in data set, $N$, and number of dry spells in the fitting
  range $\bar{N}$, and a moment ratio proportional to the cutoff are
  shown are shown. Brackets () denote errors in the last digit,
  determined by jackknife. The number of dry spells need not be within
  $\pm 1$ of the number of events, as our definition of an event (and
  a dry spell) implies that it can be split in two if it contains an
  erroneous measurement. Note the
  magnitude of this effect in the NSA data set.}  \tlabel{tdryspells}
\begin{indented}
\item[]
\begin{tabular}{l r r r r r c c } 
\br
\textbf{Site}& ${t_{d}}_{\mathrm{min}}$  & ${t_{d}}_{\mathrm{max}}$  &   ${t_{d}}_{\mathrm{max}}/{t_{d}}_{\mathrm{min}}$ &  $N$ & $\overline{N}$ & $\ave{{t_d}^2}/\ave{t_d}(er)$ & $\tau_{d}(er)$  \\ 
\mr
\footnotesize{Manus} &      24.4 &      1363.1 &        55.8 &    11992 &     4505 & 2149.(20)&   1.16(2) \\
\footnotesize{Nauru}  &     7.5 &      1027.5 &       137.7 &     5126 &     2912  &  3557.(50)&     0.99(2)\\ 
\footnotesize{Darwin}  &    8.5 &      3660.6 &       432.6 &     2892 &     1595  & 19477.(368)&    1.17(1) \\
\footnotesize{Niamey}  &   2.4 &      1774.0 &       726.1 &      262 &      135 & 26386.(1699)&    1.33(5)\\ 
\footnotesize{Heselbach} &  9.5      &      5748.0     &       605.4     &      2441 &      1035 & 2043.(34)&  1.37(2) \\
\footnotesize{Shouxian}  &  2.7     &      13488.5     &       4957.1      &      478 &      365 &  8776.(404)&    1.27(3) \\
\footnotesize{Graciosa}  & 14.6 &       415.2 &        28.5 &     3068 &     1185 & 2943.(49)  &  1.28(3)\\ 
\footnotesize{NSA}  &   12.2   &     9033.2    &     739.7    & 3440     &  1531   &  4293.(73)   &1.3(2)\\ 
\footnotesize{Pt. Reyes}          &         3.6 &     17141.0&      4826.3 &      579 &      379 &  5513.(233)          &   1.27(2)\\ 
\footnotesize{SGP }  &  8.4 &      2248.7 &       268.5 &     1625 &   523  &  17243.(463) &  1.46(3)\\
 \br
\end{tabular}
\end{indented}
\end{table}

\section{Event durations}
Precipitation event duration distributions are broad for all
locations. Durations provide a link to studies of geometric properties
of precipitation fields. Numerous studies of tropical deep convective
rain fields \cite{PetersNeelinNesbitt2009}, shallow convection fields
\cite{TrivejStevens2010}, clouds
\cite{CahalanJoseph1989,MapesHouze1993,BennerCurry1998,SchertzerLovejoy1985},
and model data from large eddy simulations
\cite{NeggersJonkerSiebesma2003} have reported the distributions of
ground covered by events (in radar snap shots etc.) to be well
approximated by power laws. We note that in the clustering null model
of critical two-dimensional percolation, clusters defined in
one-dimensional cuts, akin to durations, do not scale, whereas
two-dimensional clusters, akin to cloud-projections, do.

Applying to the durations the methods we used for the event sizes, we
find comparatively short power-law ranges, see \tref{tdurations}. The
scaling range, if it exists, is expected to be smaller than for event
sizes as the size distribution is a complicated convolution of the
event duration and precipitation rate distributions, \fref{intensities},
whose product covers a broader range than either of the distributions
alone. The event size distribution is broader than the duration
distribution also because long events tend to be more intense (not
shown).

\begin{figure}
\centering
\subfigure[]{\flabel{fdurations_a}\includegraphics*[height=0.38\textwidth]{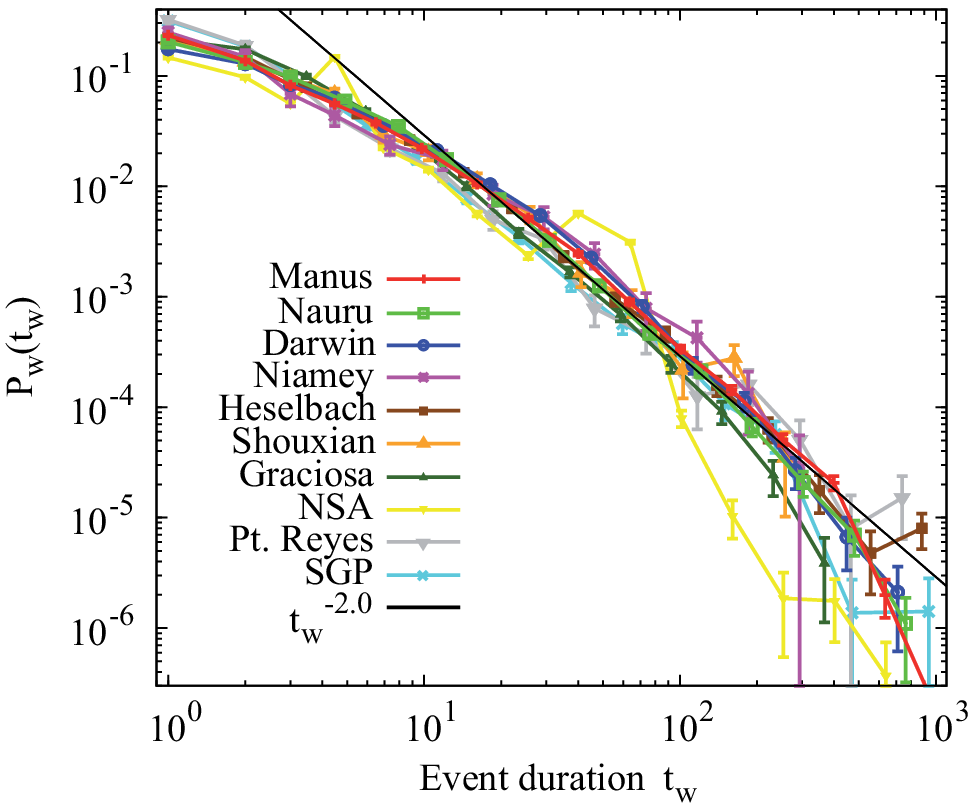}}
\hspace{0.01cm}
\subfigure[]{\flabel{fdurations_b}\includegraphics*[height=0.38\textwidth]{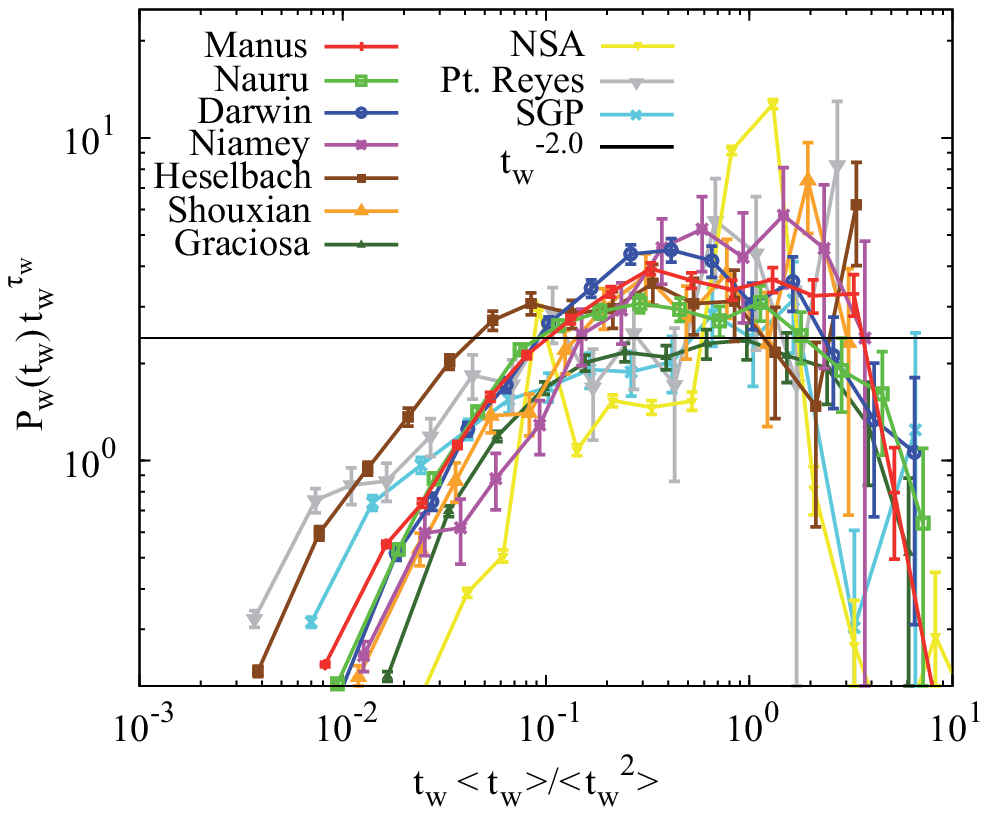}}
\caption{ (a) Probability densities for event durations (in min) are
  broad for all data sets. From a few min up to a few hundred min a
  power law with an exponent $\tau_m\approx 2.0$ roughly describes the
  data.  (b) Collapsed distributions, similar to
  \fref{scalingfunction2_b}.  } \flabel{fdurations}
\end{figure}

\begin{table}
\caption{Duration exponent (last column). Lower and upper end of
  fitting range (in min), logarithmic range
  ${t_{w}}_{\mathrm{max}}/{t_{w}}_{\mathrm{min}}$, number of events in
  data set, $N$, and number of events in the fitting range $\bar{N}$
  are shown. Brackets () denote errors in the last digit,
  determined by jackknife.}
\tlabel{tdurations}
\begin{indented}
\item[]
\begin{tabular}{ l r r r r r c c } 
\br
\textbf{Site}& ${t_{w}}_{\mathrm{min}}$  & ${t_{w}}_{\mathrm{max}}$  &   ${t_{w}}_{\mathrm{max}}/{t_{w}}_{\mathrm{min}}$ &  $N$ & $\overline{N}$ & $\ave{{t_w}^2}/\ave{t_w}$ & $\tau_{w}(er)$\\ 
\mr
\footnotesize{Manus} &        34.4 &       641.9 &        18.7 &    11981 &     1200 &  122.(1)&    2.12(4)\\
\footnotesize{Nauru}  &       25.4 &         437.5 &          17.2 &     5134 &      540 & 106.(1)& 2.09(6)\\ 
\footnotesize{Darwin}  &      17.87 &        89.30 &           5.00 &     2883 &      554 & 109.(2)& 2.0(1)\\
\footnotesize{Niamey}  &  2.7 &         211.8 &          78.4 &      262 &      157  &79.(5)&    1.39(7)\\
\footnotesize{Heselbach}  &    18.2 &   1005.0 &      55.1 &     2439 &      388 &   261.(5)&    1.97(6)\\
\footnotesize{Shouxian} &     7.7 &         197.5 &          25.5 &      480 &      172 & 84.(4) &    1.73(9)\\
\footnotesize{Graciosa}  &  12.7 &       424.0 &        33.4&     3066 &      512 &  60.(1)&  2.12(6)\\ 
\footnotesize{NSA}  &   75.2 &       103.3 &      1.4 &      9097 &      16  &  49.(1)&   6.(3)\\ 
\footnotesize{Pt. Reyes}   & 5.7 &       784.0&        138.6 &      579 &       178 &   272.(1) &  1.71(7)\\ 
\footnotesize{SGP }  &   9.4 &      278.2 &        29.7 &     1624 &      303 &  143.(4)&    1.74(7) \\
\br
 \end{tabular}
 \end{indented}
\end{table}

\section{Conclusions}
We find that the apparent avalanche size exponents, measured with
identical instruments in different locations, are consistent with a
single value of $\tau_s=1.17(3)$ for all reliable data sets. We note
that the data sets from Point Reyes and from the Southern Great Plains
are similar in many respects, despite the different reasons for
treating them with suspicion.

 The statistical error in this estimate is surprisingly
small, but neither the value itself nor the error change much using
different fitting techniques or introducing different sensitivity
thresholds (not shown). Nonetheless we believe systematic errors to be
larger. Thus, the analysis gives an impression of the universality of
the result but not necessarily the physical ``true'' value of the
exponent. This does not contradict the climatological situation --
tropical regions, for instance, are expected to support larger events
than mid-latitude locations, which could be realized as a smaller
exponent value $\tau_s$. While the exponents are not significantly
different, the larger tropical events are reflected in the greater
large-scale cutoff of the tropical distributions. Similarly, the
dry-spell durations seem to follow another power law with
$\tau_d=1.2(1)$, and regional differences can be seen in the strength
of the diurnal cycle and the cutoff dry spell duration. The broad
range of event durations, \fref{fdurations}, suggests a link to the
lack of characteristic scales in the mesoscale regime, where
approximately scale-free distributions of clusters of convective
activity, for example cloud or precipitation, have been observed to
span areas between $\mathcal{O}$(1~km$^2$) and
$\mathcal{O}$(10$^6$~km$^2$)
\cite{TrivejStevens2010,PetersNeelinNesbitt2009,NeggersJonkerSiebesma2003,BennerCurry1998,CahalanJoseph1989}.
The observation of scale-free rainfall event sizes suggests long-range
correlation in the pertinent fields, a possible indication of critical
behaviour near the transition to convective activity. Direct
measurements of the behaviour of the correlation function for the
precipitation field under changes of the (much more slowly varying)
background fields of water vapor and temperature are desirable to
clarify whether the long range correlation is a consequence of the
flow field, of the proximity to a critical point, or of a combination
of both.

\ack{}
This work was supported in part by the National Oceanic and
Atmospheric Administration Grant NA08OAR4310882 and the National
Science Foundation Grant ATM-0645200. Data were obtained from the
Atmospheric Radiation Measurement (ARM) Program sponsored by the U.S.
Department of Energy, Office of Science, Office of Biological and
Environmental Research, Environmental Sciences Division.  A. D. would
like to thank the Spanish Ministerio de Educaci\'{o}n for travel
support and Imperial College London for hospitality.  Initial research
by A.D. was supported by a grant from the Explora-Ingenio 2010 project
FIS2007-29088-E.  Other grants: FIS2009-09508, and 2009SGR-164.
A.C. is a participant of the CONSOLIDER i-MATH project.

\appendix

\section{Fitting procedure}
\label{Fitting}

In order to obtain reliable values of, for example, the exponent $\tau_s$,
independent of the binning procedure used for the plots of $P_s(s)$,
we use maximum likelihood estimation. We assume a power-law
distribution $P_s(s) = a_{\tau_s} s^{-\tau_s}$, with support
$[\smin,\smax]$. Normalization yields
$a_{\tau_s}=(1-\tau_s)/(\smax^{1-\tau_s}-\smin^{1-\tau_s}$) for a
given value of $\tau_s$.

We compute the log-likelihood function,
\begin{equation}
\mathcal{L}
:= \ln \prod_{i=1}^{\bar N}  P_s(s_i)
= \sum_{i=1}^{\bar N} \ln\left(a_{\tau_s}{s_i}^{-\tau_s}\right)
\end{equation}
where the index $i$ runs over all $\bar N$ events whose size $s_i$ 
is between $\smin$ and $\smax$.
Holding $\smin$ and $\smax$ fixed, the value of $\tau_s$ which
maximizes $\mathcal{L}$ is the maximum likelihood estimate of the
exponent. Uncertainties in $\tau_s$ are determined using the jackknife
method.

The goodness of the fit is assessed by a Kolmogorov-Smirnov (KS) test
\cite{PressETAL2002}.  The KS statistic, or KS distance, $d$, is
defined as
\begin{equation}
d :=
\displaystyle\max_{\smin \le s \le \smax}  |S_{\bar{N}}(s)-F_{s}(s)|
\end{equation}
where $S_{\bar{N}}(s)$ denotes the empirical cumulative distribution,
defined as the fraction of observed events with a size smaller than
$s$, in the interval $[\smin, \smax]$. Thus, ordering the observed
values by size, $s_1 \le \dots \le s_i \le s_{i+1}\dots \le
s_{\bar{N}}$, we have $S_{\bar{N}}(s)= i/\bar{N}$ if $s_i<s\le
s_{i+1}$; $F_{s}$ denotes the cumulative distribution of the
maximum-likelihood distribution, $F_{s}(s) := \int_{\smin}^s P_s(t)
dt$.

The KS distance translates into the $p-$value. The $p-$value is the
probability that synthetic data, here drawn from a power law
distribution with exponent $\tau_s$, result in a KS-distance of at
least $d$. For instance, $p=10\%$ means that for power-law distributed
data with exponent $\tau_s$ there is a probability of 0.90 that the KS
distance takes a value smaller than $d$. Thus, if the data really are
generated by a power law and we decide to reject the power law as a
model if $p<10\%$, we will reject the correct model in 10\% of our
tests.  Conversely, decreasing the limit of rejection in the $p-$value
implies that we accept more false models.

In our implementation of the KS test the distribution to be tested,
$P_s(s)$, is not independent of the empirical data. This is because
the exponent $\tau_s$ is obtained from the data that are later used to
test the distribution. We therefore cannot use the standard analytic
expression for $p(d)$, see Ref.~\cite{PressETAL2002}, Ch.~15.
Instead, we determine the distribution of the KS distance and
therefore the $p-$value by means of Monte Carlo simulations: we
generate synthetic power-law-distributed data sets between $\smin$ and
$\smax$ with exponent $\tau_s$ and number of data $\bar N$ (see
Table~\ref{table: maxnumber10}), and proceed exactly in the same way
as for the empirical data, first obtaining a maximum likelihood
estimate of the exponent $\tau_s$ and then computing the KS distance
between the empirical distribution of the simulated data and the
fitted distribution containing the estimated value of $\tau_s$. The
$p-$value is obtained as the fraction of synthetic data sets for which
the KS statistic is larger than the value obtained for the empirical
data.

The final step is to compare results for different ranges $[\smin,
  \smax]$. We try all possible fitting ranges with $\smin$ and $\smax$
increasing by factors of $10^{0.01}\approx 1.023$. We choose to report
those intervals $[\smin, \smax]$ that contain the largest number of
events $\bar N$ with a corresponding $p-$value larger than 10\%.

\section{Two-sample Kolmogorov-Smirnov Tests}
\label{Two-sample}
A two-sample Kolmogorov-Smirnov test was performed for each pair of
data sets, $i, j$ to test whether the two underlying event-size
probability distributions differ. This test does not assume any
functional form for the probability distributions
\cite{PressETAL2002}. As in the fitting of the exponent, we vary the
testing ranges $[\smin, \smax]$, keeping those which yield $p > 10
\%$. We report the range with the maximum effective number of data,
$\bar N_{\mathrm{eff}}\equiv \bar{N_i} \bar{N_j} /(\bar{N_i} +
\bar{N_j})$. The results, shown in Table~\ref{table: varkstest2a2},
confirm that the pairs of distributions from the reliable data sets
are similar over broad ranges.

\begin{table}
  \caption{ 
Maximum range $\smax/\smin$ over which the $p$-value of
of a two-sample KS test is greater than 10\%.
   \label{table: varkstest2a2}}
\begin{tabular}{c c c c c c c c c c } 
\br
  \phantom{a}& \footnotesize{Nauru} & \footnotesize{Darwin} & \footnotesize{Niamey} & \footnotesize{Heselbach} & \footnotesize{Shouxian} & \footnotesize{Graciosa} & \footnotesize{NSA} & \footnotesize{Pt. Reyes} & \footnotesize{SGP} \\ 
 \mr 
 \footnotesize{Manus}     &  \footnotesize{5386.} &   \footnotesize{16257.}  &   \footnotesize{16386.} &    \footnotesize{679.}   &    \footnotesize{6355.}   & \footnotesize{638.} &\footnotesize{14.} &\footnotesize{ 32.} & \footnotesize{8.} \\
 \footnotesize{Nauru}     &   -     &    \footnotesize{6753.}  &   \footnotesize{13495.} &    \footnotesize{236.}   &    \footnotesize{221.}    & \footnotesize{342.}& \footnotesize{27.}& \footnotesize{19.}& \footnotesize{7.} \\
 \footnotesize{Darwin}    &   -     &    -     &   \footnotesize{12247.} &    \footnotesize{236.}   &    \footnotesize{271.}    & \footnotesize{575.} & \footnotesize{27.} &\footnotesize{16. }  &  \footnotesize{5.}   \\
 \footnotesize{Niamey}    &   -     &    -     &    -    &   \footnotesize{3466.}   &    \footnotesize{16420.}  &  \footnotesize{2358.}&  \footnotesize{  1599.}&  \footnotesize{668.}& \footnotesize{253.} \\
 \footnotesize{Heselbach} &   -     &   -      &    -    &     -    &    \footnotesize{14600.}  &  \footnotesize{13265. }&  \footnotesize{   18.}& \footnotesize{20.} &  \footnotesize{5.} \\
 \footnotesize{Shouxian}  &  -     &   -      &    -    &     -    &    -   &   \footnotesize{26440.}&  \footnotesize{13.}& \footnotesize{65.} &   \footnotesize{39.} \\
 \footnotesize{Graciosa}  &  -     &   -      &    -    &     -    &    -    &- &  \footnotesize{11.}&  \footnotesize{ 17.}&   \footnotesize{589.} \\
 \footnotesize{NSA}        &  -     &   -      &    -    &     -    &   -  & -&- &  \footnotesize{10.}&\footnotesize{ 3.} \\
 \footnotesize{Pt. Reyes}&  -     &   -      &    -    &     -    &    -   & -&- &- &\footnotesize{19916.} \\ 
                
 \br
 \end{tabular}
\end{table}

\begin{verbatim*}

\end{verbatim*}

\section*{References}
\bibliographystyle{unsrt}

\bibliography{20100928}

\end{document}